\documentclass[aps,prl,preprint,groupedaddress]{revtex4}
\usepackage{graphicx}
\begin{document}

\title{Hollowgraphy Driven Holography:\\
Black Hole with Vanishing Volume Interior\footnote{Honorable
Mention Award - Gravity Research Foundation (2010)}}

\author{Aharon Davidson and Ilya Gurwich}

\affiliation{Physics Department, Ben-Gurion University,
Beer-Sheva 84105, Israel\\
Email: davidson@bgu.ac.il, gurwichphys@gmail.com}

\date{April 1, 2010; July 16, 2010}

\begin{abstract}
Hawking-Bekenstein entropy formula seems to tell us that no quantum
degrees of freedom can reside in the interior of a black hole.
We suggest that this is a consequence of the fact that the volume
of any interior sphere of finite surface area simply vanishes.
Obviously, this is not the case in general relativity.
However, we show that such a phenomenon does occur in various
gravitational theories which admit a spontaneously induced
general relativity.
In such theories, due to a phase transition (one parameter family
degenerates) which takes place precisely at the would have been horizon,
the recovered exterior Schwarzschild solution connects, by means
of a self-similar transition profile, with a novel 'hollow' interior
exhibiting a vanishing spatial volume and a locally varying Newton
constant.
This constitutes the so-called 'hollowgraphy' driven holography.
\end{abstract}

\maketitle

The Hawking-Bekenstein\cite{BHentropy} black hole entropy formula
\begin{equation}
	S_{BH}=\frac{c^{3}A_{BH}}{4G\hbar}
	\label{SBH}
\end{equation}
constitutes a triple point in the phase space of physical theories.
This formula touches gravity, even beyond general relativity (GR),
quantum mechanics, and statistical mechanics; and thus, is
expected to play a major role in (the still at large) quantum
gravity.
In contrast with ordinary macroscopic systems, whose
entropy is known to be proportional to their volume $V$, the
entropy of a black hole is intriguingly proportional to the surface
area $A_{BH}$ of the event horizon.
Following conventional statistical mechanics wisdom, this seems
to tell us that, from some fundamental as yet obscure reason, no
degrees of freedom can reside within the interior of a black hole.
Recall that the Gibbons-Hawking\cite{GibbonsHawking} derivation
of eq.(\ref{SBH}) does not make use of the black hole interior.
Also Wald's\cite{Wald} derivation, being more locally oriented,
solely invokes geometrical properties of the horizon itself, but
without addressing the interior regime.

The area entropy has inspired the holographic
principle\cite{Hprinciple}.
The latter, primarily introduced by 't Hooft\cite{tHooft} to
resolve the black hole information paradox, has been further developed
by Susskind\cite{Susskind}, gaining theoretical support from the
AdS/CFT duality\cite{AdsCFT}.
However, in spite of the impressive theoretical progress, a basic
question is still to be answered.
Namely, why is the black hole entropy proportional to the area of
the black hole horizon rather than to its volume?
Unfortunately, GR per se does not provide an explanation.
Given the Schwarzschild solution, and owing to the
$r\leftrightarrow t$ signature flip, the volume of any sphere
of a finite surface area is problematic\cite{Vproblem}.
But even if $V_{BH}$ would have been well behaved, then
following a conventional  asymptotic expansion\cite{V+A}
$S_{BH}=\xi V_{BH}+\eta A_{BH} +...$, either the coefficient
$\xi$, or $V_{BH}$ itself, must vanish for the area term to dominate
the entropy.
Unfortunately, as far as GR is concerned, the exterior
Schwarzschild solution 'refuses' to connect with an interior core of
zero volume.
However, if GR is not the final word, and is either
(i) A limit of a more sophisticated theory of gravity, or even
(ii) Just supplemented by various field theoretical corrections,
things can change drastically.
In this paper we raise the intriguing possibility that the surface
contribution to the black hole entropy prevails simply because
any inner sphere, while being characterized by a finite surface
area $A(r)=4\pi r^{2}$, happens to exhibit a vanishing volume
$V(r)\rightarrow 0$.
We refer to such a idea as 'Hollowgraphy' driven Holography.

Serendipitously, the realization of such an idea only calls for a
slight deviation from GR, parametrized by some small
parameter $\epsilon$.
Whereas the exact Schwarzschild solution is fully recovered for
$\epsilon =0$, the 1-parameter family of solutions degenerates
(this is also known as 'level crossing') as $\epsilon\rightarrow 0$,
precisely on the would have been event horizon, with the recovered
exterior Schwarzschild solution being analytically connected now
with a novel interior core.
Horizon phase transition has already been discussed in the
literature\cite{PhaseTransition}, and so was the idea that quantum
effects may prevent black holes from forming, and instead
give rise to black stars\cite{BlackStar}.
However, a $V_{BH}\rightarrow 0$ black hole/star configuration
has never been demonstrated.

To concretize the idea, let us first add a simple curvature quadratic to
the Einstein-Hilbert action
\begin{equation}
	I=\int \left(-\frac{{\cal{R}}}{16\pi G}+
	\frac{a}{16\pi}{\cal{R}}^{2}+{\cal L}_{m}\right)
	\sqrt{-g}~d^{4}x ~.
	\label{R+R2}
\end{equation}
Note that such a term can arise from quantum corrections\cite{BirrellDavies}.
While facing GR for $a=0$, it is incorrect to assume that adding
higher derivative correction terms, e.g. $f(R)$ gravity\cite{f(R)},
with 'small' coefficients will only produce
small modifications of the solutions of the unperturbed theory.
To provide extra insight, notice that the above action is
equivalent\cite{equivalence} to a certain $\omega=0$ (or
$\omega=3/2$ using the Palatini formalism) Brans-Dicke (BD)
theory\cite{BDgravity} supplemented by a quartic Higgs type potential
\begin{equation}
	I=\int \left(-\frac{\varphi^{2}{\cal{R}}}{16\pi }-W(\varphi)+
	{\cal L}_{m}\right)
	\sqrt{-g}~d^{4}x ~,
	\label{BD}
\end{equation}
with
\begin{equation}
	W(\varphi)=\frac{1}{64\pi a}\left( \varphi^{2}-G^{-1}\right)^{2}~.
\end{equation}
We remark that (i) An arbitrary $\omega$ BD scalar kinetic term can be added
without affecting our main conclusion, and that (ii) The scalar field
$\varphi(x)$ is dynamical even in the absence of such a kinetic term.
The role of the potential is to allow, by virtue of the Zee\cite{Zee}
mechanism, for the spontaneous emergence of GR
once the BD scalar approaches its VEV, that is
$\langle \varphi \rangle=G^{-1/2}$.
The fact that the matter Lagrangian ${\cal L}_{m}$ does not couple
to the BD scalar elevates the Jordan frame, rather than the Einstein
frame, to the level of the physical frame.

Associated with a static spherically symmetric line element
\begin{equation}
	ds^{2}=-e^{\nu(r)}dt^{2}+
	e^{\lambda(r)}dr^{2}+r^{2}d\Omega^{2}~,
\end{equation}
are  the field equations (using the notation $\phi\equiv\varphi^{2}$)
\begin{eqnarray} \nonumber
	&&\phi^{\prime\prime}-\frac{1}{2}(\nu^{\prime}+\lambda^{\prime})
	\left(\phi^{\prime}+\frac{2}{r}\phi\right)=0 ~,\\
	&&\phi^{\prime\prime}+\frac{1}{2}(\nu^{\prime}-\lambda^{\prime})
	\left(\phi^{\prime}-\frac{2}{r}\phi\right)-\frac{2}{r^{2}}(1-e^{\lambda})
	\phi =\frac{2}{3}e^{\lambda}
	\left(\phi\frac{dW(\phi)}{d\phi}-\frac{1}{2}W(\phi)\right) ~,
	\\ \nonumber
	&&\phi^{\prime\prime}+
	\left(\frac{\nu^{\prime}-\lambda^{\prime}}{2}+
	\frac{2}{r}\right)\phi^{\prime}=\frac{1}{3}e^{\lambda}
	\left(\phi\frac{dW(\phi)}{d\phi}-2W(\phi)\right) ~.
	\label{FieldEquations}
\end{eqnarray}
The scalar field equation
$\displaystyle{R+\frac{1}{2a}\left( \phi-\frac{1}{G}\right)=0}$,
which is a certain combination of the above equations,
will play an essential role in regions where the Newton constant
is away from its VEV.
The asymptotically flat, large distance (and small mass) expansion
is quite conventional, with the scalar charge $\epsilon$ factorizing
a variety of Yukawa suppressed terms at a typical length scale
$\Omega=\sqrt{a G}$, namely
\begin{eqnarray}\nonumber
	&& \phi(r) \simeq \frac{1}{G}\left(
	1+ \epsilon\frac{e^{-2r/\Omega}}{r^{1+2GM/\Omega}}F(r)\right)~,\\
	&& e^{\nu(r)}\simeq
	1-\frac{2GM}{r}+\frac{2\epsilon}{\Omega}
	\frac{e^{-2r/\Omega}}{r^{2GM/\Omega}}N(r)~,
	\\  \nonumber
	&& e^{-\lambda(r)}\simeq
	1-\frac{2GM}{r}+\frac{2\epsilon}{\Omega}
	\frac{e^{-2r/\Omega}}{r^{2GM/\Omega}}L(r)~,
	\label{}
\end{eqnarray}
where the various functions involved exhibit the long distance behavior 
\begin{equation}
	F(r),~N(r),~ L(r) = 1 +{\cal O}(r^{-1}) ~.
\end{equation}

\begin{figure}[ht]
	\includegraphics[scale=0.95]{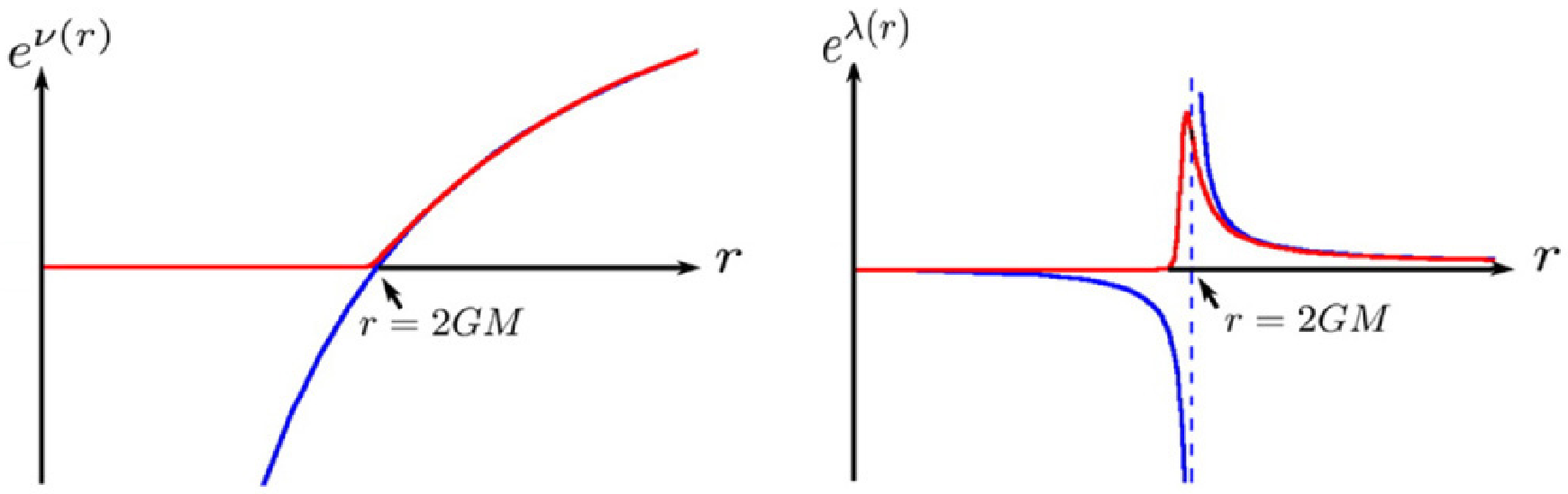}
	\includegraphics[scale=0.95]{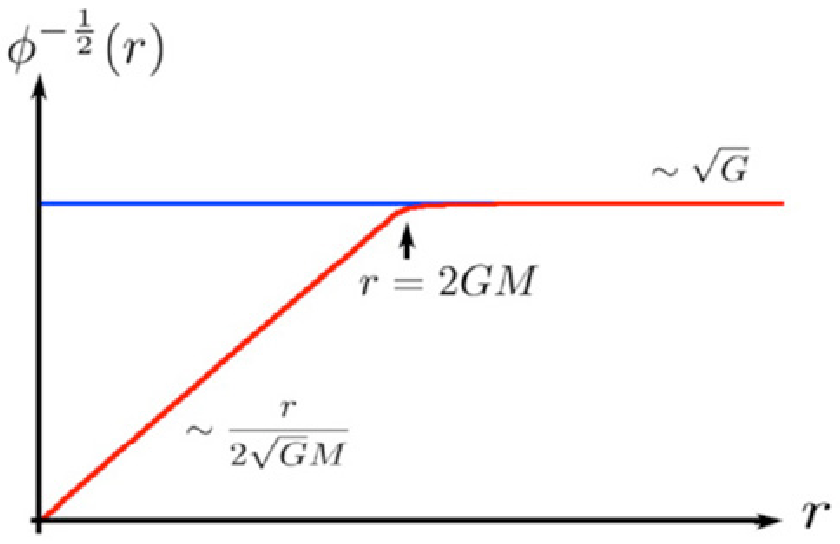}
	\caption{The components of the static spherically symmetric
	solution are plotted as a function of the radial marker $r$.
	The $\epsilon\ne 0$ solution (red) vs. the $\epsilon=0$
	Schwarzschild solution (blue).
	At the limit $\epsilon\rightarrow +0,$
	a phase transition occurs on the would have been horizon
	(at $r\rightarrow 2GM$).}
	\label{Fig1}
\end{figure}

The Schwarzschild solution is fully recovered, not just asymptotically,
for $\epsilon=0$.
However, with the focus on the $\epsilon\rightarrow +0$ limit, one
can numerically run a full scale solution, see Fig.\ref{Fig1}, and
already suspect the emergence of a phase transition at the would have
been horizon.
While the exterior Schwarzschild solution is asymptotically recovered,
it is clear that as far as the inner core is concerned,
(i) The $t \leftrightarrow r$
signature flip is gone, (ii) $e^{\nu(r),\lambda(r)}$ get drastically
suppressed, and (iii) The effective Newton constant $\varphi^{-1}$
ceases to be constant.

To isolate the new geometrical branch which enters the game, we
momentarily neglect the $e^{\lambda}(r)$ terms in the field equations
(the consistency of this approximation is later verified) to face the exact
short distance analytic behavior
\begin{equation}
	e^{\lambda (r)}\simeq\beta
	\left(\frac{r}{2GM}\right)^{2\left(\frac{3}{\delta}-3+\delta\right)} ~,\quad
	e^{\nu (r)}\simeq\alpha
	\left(\frac{r}{2GM}\right)^{2\left(\frac{3}{\delta }-2\right)} ~,\quad
	\phi (r)\simeq \frac{1}{G}  \left(\frac{r}{2GM}\right)^{-2+\delta }~,
	\label{shortdistance}
\end{equation}
with the emerging constant of integration $\delta$ being apparently arbitrary
at this stage.
However, noticing that $R\simeq-2/r^{2}$, and owing to the $R\sim \phi$
behavior at short distances, it becomes obvious that once the $e^{\lambda}(r)$
terms are to be re-introduced, then it is $\delta \rightarrow +0$ which
becomes relevant fo the black hole/star.
The exact relation between $\delta$ and $\epsilon$ is generally quite
complicated, and at this stage, was only obtained numerically
by plotting $\frac{1}{2}r(\lambda^{\prime}+\nu^{\prime})$.
However, for the limiting case of a negligible scalar
potential (large $\Omega$), we find
\begin{equation}
	\delta \simeq \frac{3\epsilon}{2GM}~.
\end{equation}
Whereas this accounts for the small $r$ regime, it still lacks the physical
scale where this behavior actually terminates to connect with the exterior
branch (where $\phi(r)$ approaches the
VEV set by the potential).

To analytically derive the transition profile, we adopt the following
technique.
We only keep in the equations those terms
which vary drastically at the neighborhood of the transition.
In particular, it turns out that the transition profile is not sensitive to
the terms involving the scalar potential $W(\varphi)$.
One equation of motion can then be integrated directly, leading to the
conserved charge
$C=-\phi^{\prime}(r)r^{2}e^{\frac{1}{2}\left(\nu(r)-\lambda(r)\right)}$.
This paves the way for the transitionary solution
\begin{eqnarray} \nonumber
	&& e^{\lambda (r)}\simeq p e^{\sigma (r)}\left(\frac{4M}{C}-
	e^{\sigma (r)}\right) ~, \\
	&& e^{\nu (r)}\simeq p e^{-\sigma (r)}\left(\frac{4M}{C}-
	e^{\sigma(r)}\right) ~,\\
	&& r-\bar{r}\simeq \frac{GC}{2p}\left(e^{-\sigma (r)}+
	\frac{C }{4M}\log\left(\frac{4M}{C} e^{-\sigma (r)}-1\right)\right) ~,
	\nonumber
\end{eqnarray}
with $e^{\sigma (r)}$ serving as a parametric function, and $\bar{r},p$
as constants of integration.
This solution allows for matching with both
the exterior Schwarzschild near horizon behavior as well as the interior
solution specified earlier.
This allows us to fix $\bar{r}=2GM$, and to express the coefficients
$\alpha,\beta$ and also $p,C$ in terms of $\delta$, namely
\begin{equation}
	\alpha=\frac{\delta}{6}~,\quad
	\beta=\frac{6}{\delta}~,\quad
	p=\frac{C}{4M}=\frac{\delta}{6}~.
\end{equation}
The above solution admits a remarkable self-similar structure.
Namely, $\delta\rightarrow k\delta$
only causes scale changes $e^{\lambda}\rightarrow k^{-1}e^{\lambda}$
and $r-2GM\rightarrow k(r-2GM)$.
The emerging transition profile then resembles, in some sense, the
stretched horizon\cite{stretched} which characterizes the brane paradigm.

Altogether, the inner metric is well approximated  by
\begin{equation}
	ds^{2} \simeq -\frac{\delta}{6}
	\left(\frac{r}{2GM}\right)^{2\left(\frac{3}{\delta}-
	2\right)}dt^{2}+
	\frac{6}{\delta}\left(\frac{r}{2GM}\right)^{2\left(\frac{3}{\delta}-
	3\right)}dr^{2}+r^{2}d\Omega^{2} ~.
\end{equation}
Notice that the effective Newton constant ceases to
be a constant in the inner core
\begin{equation}
	G_{in}(r) \simeq  G \left(\frac{r}{2GM} \right)^{2-\delta} ~.
\end{equation}
The fingerprint of the above metric is the invariant volume $V(r)$,
associated with an inner sphere of a finite surface area $4\pi r^{2}$,
given by
\begin{equation}
	V(r)\ \simeq 4\pi \sqrt{\frac{2\delta}{3}}(2GM)^{3}
	\left(\frac{r}{2GM}\right)^{3/\delta},
	\quad \text{for}~ r\leq 2GM ~.
\end{equation}
\begin{figure}[ht]
	\includegraphics[scale=1]{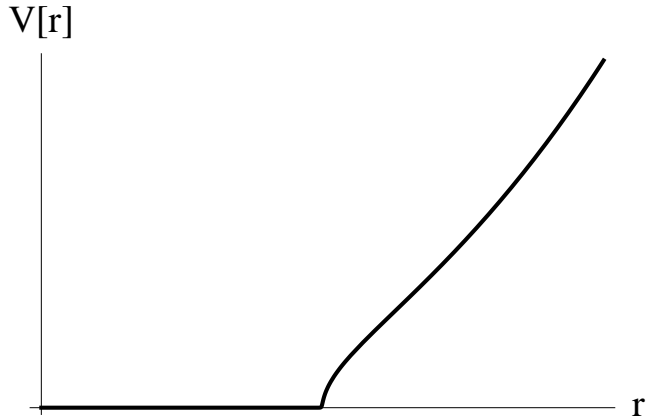}
	\caption{The invariant volume $V(r)$ associated with
	a sphere of a surface area $4\pi r^{2}$.
	The $V_{BH}\rightarrow 0$ property can explain
	why no degrees of freedom reside inside a black hole/star.}
	\label{Fig2}
\end{figure}
For $\delta>0$, $V(r)$ is a smooth monotonically growing function
of $r$.
For $\delta\rightarrow +0$, $V(r)$ vanishes inside, and starting
from the would have been horizon, it grows until asymptotically
approaching the $\frac{4}{3}\pi r^{3}$ value.

To gain insight at the neighborhood of the origin, one may re-define
the radial coordinate
\begin{equation}
	\rho\simeq 2GM \sqrt{\frac{2\delta}{3}}
	\left(\frac{r}{2GM}\right)^{\frac{3}{\delta}-2}~,
\end{equation}
and write the inner metric in the form
\begin{equation}
	ds^2\simeq-\left(\frac{\rho}{4GM}\right)^2dt^2+d\rho^2+
	4G^{2}M^{2}\left(\frac{3\rho^{2}}{8G^{2}M^{2}\delta}
	\right)^{\frac{\delta}{3}}
	d\Omega^2.
\end{equation}
The Rindler structure of the temporal-radial part of the metric
implies that the singularity in the origin is now protected by
a 'pseudo-horizon'.
The imaginary time periodicity needed to avoid the associated conic
singularity then implies, quite unexpectedly, that this pseudo-horizon
exhibits the exact Hawking temperature as the Schwarzschild horizon.

The emerging picture is highly non-standard yet very pleasing.
From the outside, an observer cannot tell the 'hollow' black hole
geometry from the Schwarzschild spacetime.
However, due to a novel phase transition
which takes place precisely at the would have been horizon, the recovered exterior
Schwarzschild solution connects, by means of a special self-similar
transition profile, with a novel 'hollow' interior.
It is the $V_{BH}\rightarrow 0$ property which can then account
for the fact that no degrees of freedom can reside inside a black hole
(so that the black hole entropy is dominated by the finite $A_{BH}\neq 0$),
thereby constituting the so-called 'hollowgraphy' driven holography.

\acknowledgments{It is our pleasure to cordially thank  Shimon Rubin
for his support and valuable remarks, and Eduardo Guendelman for
inspiring discussions.}

\end{document}